%% file: main.tex
\documentclass[letterpaper]{article} 

\usepackage{aaai2026}

\usepackage{times}  
\usepackage{helvet}  
\usepackage{courier}  
\usepackage[hyphens]{url}  
\usepackage{graphicx} 
\usepackage{natbib}  
\usepackage{caption} 
\usepackage{algorithm}
\usepackage{algorithmic}

\usepackage{xspace}
\newcommand{\themodel}{DugFND\xspace}
\usepackage{amsmath}
\usepackage{booktabs}
\usepackage{adjustbox}  
\usepackage[many]{tcolorbox}
\usepackage{multirow}
\usepackage{colortbl}
\usepackage{tabularx}
\usepackage{xcolor}

\usepackage{newfloat}
\usepackage{listings}
\DeclareCaptionStyle{ruled}{labelfont=normalfont,labelsep=colon,strut=off} 
\floatstyle{ruled}
\newfloat{listing}{tb}{lst}{}
\floatname{listing}{Listing}
\title{Mining the Social Fabric: Unveiling Communities for Fake News \\ Detection in Short Videos}
\author {
    Haisong Gong\textsuperscript{\rm 1,\rm 2},
    Bolan Su\textsuperscript{\rm 3},
    Xinrong Zhang\textsuperscript{\rm 3},
    Jing Li\textsuperscript{\rm 1,\rm 2},\\
    Qiang Liu\textsuperscript{\rm 1,\rm 2},
    Shu Wu\textsuperscript{\rm 1,\rm 2},
    Liang Wang\textsuperscript{\rm 1,\rm 2}
}
\affiliations{
\textsuperscript{\rm 1}New Laboratory of Pattern Recognition (NLPR), Institute of Automation, Chinese Academy of Sciences\\
\textsuperscript{\rm 2}School of Artificial Intelligence, University of Chinese Academy of Sciences\\
\textsuperscript{\rm 3}Douyin Group

}

\nocopyright

\begin{document}

\maketitle

\begin{abstract}
Short video platforms have become a major medium for information sharing, but their rapid content generation and algorithmic amplification also enable the widespread dissemination of fake news. Detecting misinformation in short videos is challenging due to their multi-modal nature and the limited context of individual videos. While recent methods focus on analyzing content signals—visual, textual, and audio—they often overlook implicit relationships among videos, uploaders, and events. To address this gap, we propose \textbf{\themodel} (\underline{Du}al-community \underline{g}raph for \underline{f}ake \underline{n}ews \underline{d}etection), a novel method that enhances existing video classifiers by modeling two key community patterns: (1) uploader communities, where uploaders with shared interests or similar content creation patterns group together, and (2) event-driven communities, where videos related to the same or semantically similar public events form localized clusters. We construct a heterogeneous graph connecting uploader, video, and event nodes, and design a time-aware heterogeneous graph attention network to enable effective message passing. A reconstruction-based pretraining phase further improves node representation learning. \themodel can be applied to any pre-trained classifier. Experiments on public datasets show that our method achieves significant performance gains, demonstrating the value of dual-community modeling for fake news detection in short videos.
\end{abstract}


\input{introduction}
\input{relatedwork}

\input{method}

\input{experiment}

\input{conclusion}

\bibliography{aaai2026}

\end{document}

%% file: introduction.tex
\section{Introduction}
Short videos have emerged as a dominant form of information dissemination, empowering users to create and share content at scale through algorithmic recommendation systems~\cite{chen2022tiktok,violot2024shorts}. However, the daily generation of hundreds of millions of short videos poses significant cybersecurity risks: malicious actors craft multi-modal deceptive content to artificially inflate traffic for profit, eroding social trust and destabilizing online ecosystems~\cite{aimeur2023fake}. As short video platforms expand globally, developing robust algorithms to effectively identify and detect malicious fake videos has become critically important.

Fake news detection in the short video domain involves distinguishing between fake and real content~\cite{shu2017fake}, but this task is complicated by the unique challenges posed by the multi-modal nature of the videos. Unlike traditional methods that primarily focus on textual content and comments~\cite{zhang2024breaking,liu2024out}, short videos integrate heterogeneous information sources such as uploader profiles, user interaction patterns, and multi-modal content (e.g., video, text, audio), which together introduce significant complexity in detecting fake news.

At the current stage, many methods have been developed for detecting fake news in short videos. Most of these approaches primarily focus on analyzing the video content itself: integrating visual, textual, and audio features~\cite{ren2024mmsfd}; detecting inconsistencies across multi-modal messages~\cite{bu2024fakingrecipe}; constructing powerful encoders~\cite{wu2024interpretable}; or leveraging large language models to mine implicit opinions of video uploaders~\cite{zong2024unveiling}. These methods have made great progress in detecting fake signals in individual videos, but they neglect the rich information embedded in the relationships between videos. We term these existing approaches as individual video classifiers. Recent works~\cite{li2025learning, qi2023two} attempt to bridge this gap using graph neural networks, but they construct homogeneous graphs limited to intra-event video links. Crucially, they overlook the broader structures that span across events and uploaders, where implicit but informative communities often emerge. These communities—shaped by common interests, shared content patterns, or topics—hold key signals for understanding how misinformation circulates at scale, yet remain largely underexplored.

The implicit communities at play in misinformation detection unfold in two key dimensions: (1)\textit{``Birds of a feather flock together.''} Video uploaders often operate within implicit communities shaped by shared interests, similar content creation patterns, and, at times, malicious intent. These communities naturally emerge through user interactions on the platform, where individuals gravitate toward others with aligned views or objectives~\cite{cinelli2021echo}. In the context of misinformation, such clustering behavior becomes particularly problematic, as uploaders within these communities often collaborate or mimic each other's content, amplifying deceptive narratives. (2) \textit{``Where there’s smoke, there’s fire.''}  Public events tend to attract a surge of content, both legitimate and fraudulent, and videos related to these events form their own communities. These event-driven communities exhibit distinct propagation patterns~\cite{liu2015events}, reflecting unique linguistic and stylistic features~\cite{francis2024variation}. For example, during health-related events, content may exaggerate facts, while in political contexts, misinformation may focus on creating division or confusion. Recognizing these patterns of uploader clustering and event-driven content formation is key to identifying coordinated misinformation campaigns, enabling more effective detection of fake news spread across multiple videos.

Based on the above analysis, we propose \textbf{\themodel} (\underline{Du}al-community \underline{g}raph for \underline{f}ake \underline{n}ews \underline{d}etection), a novel method that integrates dual-community patterns—uploaders' communities and event-driven communities—to enhance any pre-trained individual video classifier. Specifically, we construct a heterogeneous graph encompassing three node types—uploaders, videos, and events—to capture latent associations. This graph is built through two key processes: forming uploader communities by grouping uploaders with similar profiles, and constructing event communities via intra-event co-occurrence relations and inter-event semantic connections. Video features extracted from a base classifier are then enriched through message passing over this graph. To guide information flow effectively, we design a heterogeneous graph attention network with a time-aware mechanism that models the temporal dynamics across videos, events, and their interactions. In addition, we introduce a reconstruction-based pretraining phase to help the model learn transferable node representations before final fine-tuning for video classification. \themodel requires no external data, and generalizes well across datasets. Experiments on public datasets demonstrate substantial performance gains across multiple base models, validating the benefits of dual-community modeling for fake news detection in short videos.

As a summary, our major contributions are:
\begin{itemize}
    \item We highlight the limitations of traditional methods, which ignores the rich community relationship between videos, uploaders and events.
    \item We propose \themodel, a method that applies dual-community insights to an individual video classifier, improving its ability to detect fake news.
    \item Experiments show that \themodel significantly improves performance across public benchmarks, validating the effectiveness of integrating dual-community.

\end{itemize}

%% file: relatedwork.tex
\section{Related Works}

\subsection{Multi-Modal Fake News Detection}
Early fake news detection methods primarily focused on textual content, often analyzing linguistic features such as sentiment~\cite{ercsahin2017twitter} or leveraging user comments from social platforms~\cite{zhang2024breaking}. In contrast, multi-modal fake news detection presents a more complex challenge, as it requires reasoning over heterogeneous signals including video, audio, and textual metadata. In the context of short videos on social platforms, numerous recent studies have explored this task.
SVFEND~\cite{qi2023fakesv} proposes a unified architecture that integrates features from titles, descriptions, video frames, audio, comments, and uploader profiles. FANVM~\cite{choi2021using} employs adversarial learning and topic modeling to learn topic-agnostic representations. FakingRecipe~\cite{bu2024fakingrecipe} designs targeted modules to detect modality inconsistency and manipulation artifacts in edited videos. MMVD~\cite{zeng2024mitigating} addresses cognitive biases by multi-view causal reasoning strategy, while MMSFD~\cite{ren2024mmsfd} introduces a multi-grained fusion strategy to enhance cross-modal alignment.
Recently, large language models (LLMs) have gained popularity in detecting misinformation~\cite{gong2024navigating,gong2025strive}. OpEvFake~\cite{zong2024unveiling} and VMID~\cite{zhong2024vmid} use LLMs to generate explanatory analyses that are then incorporated as features for classification. ExMRD~\cite{hong2025following} leverages LLMs with chain-of-thought prompting and fine-tuning to support fake short video detection. While these methods show strong performance, they treat each video independently, without modeling inter-video relationships.

\subsection{Graph Methods for Fake News Detection}
 Graph-based methods have proven effective for modeling complex relational structures~\cite{zhao2017meta,gong2024heterogeneous}. In traditional fake news detection, graphs are often constructed over news propagation structures, where nodes represent news posts and edges represent interactions such as reposts or comments~\cite{bian2020rumor,zhang2024gbca,zhu2024propagation}. Recent approaches have enriched these graphs with semantic nodes derived from topics or entities via LLMs~\cite{ma2024fake}, enabling deeper reasoning. In the short video domain, graph-based methods are still emerging. \citet{qi2023two} and \citet{li2025learning} represent individual videos as nodes and construct intra-event graphs to enhance misinformation detection. However, these approaches primarily focus on video-level relations within a single event. In contrast, our work expands the graph to include both uploader and event communities, capturing dual-community structures that encode broader social and semantic connections across videos.

%% file: method.tex
\newtcolorbox{mybox}{
    enhanced,
    breakable,
    boxrule = 0.47pt,
    colback = sub,
    left = 0mm,
    right = 0mm,
    top = 1mm,
    bottom = 1mm,
    before skip = -2mm,
    after skip = 2mm,
    arc = 0pt   
}
\definecolor{sub}{HTML}{ffffff}

\section{Methodology}
\subsection{Problem Formulation}
Given dataset $D=D_\text{labeled}\cup D_\text{unlabeled}$, the core task of fake news detection in short videos is to identify fake news videos within $D_\text{unlabeled}$. Each video item $x_i\in D$ contains multi-modal information, expressed as: 
\begin{equation}
x_i=\{x_i^f,x_i^t,x_i^c,x_i^a, x_i^u,x_i^e\}     
\end{equation}
where $x_i^f,x_i^t,x_i^c,x_i^a$ refers to frame information (e.g. video frames, scenes), video textual content (e.g. title, ocr, asr), video comments and audio, respectively. Additionally, $x_i^u$ and $x_i^e$  refers to the uploader’s profile information and the event-related information connected to the video, which serve as the foundation for the implicit community structure in \themodel. The goal is to develop a model that predicts $\hat p\in\{0,1\}$ for each video in $D_\text{unlabeled}$, determine whether the video is fake or not.

\begin{figure*}
    \centering
    \includegraphics[width=\textwidth]{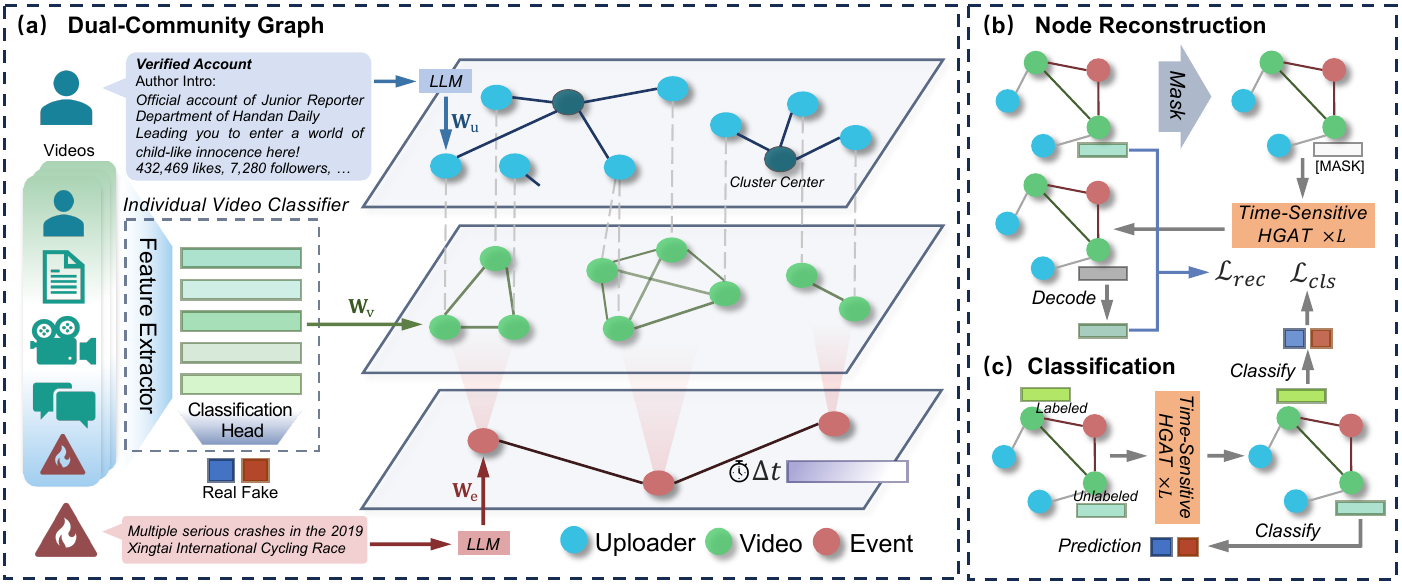}
    \caption{(a) Dual-community graph construction; (b) Masked node pretraining; (c) Supervised fine-tuning and inference.}
    \label{fig:main}
\end{figure*}

\subsection{Multi-Modal Feature Extraction}
Previous methods have made significant efforts in integrating multi-modal information from the uploaded video content. For example, one high-performance models, SVFEND employs a cross-attention mechanism to exploit cross-modal correlations, while also utilizing social context information for detection.

Our approach, by contrast, avoids adding complexity to multi-modal fusion. Instead, we leverages an individual video classifier $F$ to extract features from each video $x_i$, using part or all of the available information in $x_i$. Rather than relying on the classifier for direct prediction, we focus on the feature representations from the layer preceding the classification head. These features are used to construct the encoded content representation of the video $x_i$, denoted as $\textbf{x}_{\textbf{v},i}^0$:
\begin{equation}
    \textbf{x}_{\textbf{v},i}^{0} = \mathbf{W}_vF(x_i)\quad \forall x_i\in D
\end{equation}
where $\mathbf{W}_v$ represents a learnable projection matrix.

\subsection{Dual-Community Graph Construction}
In \themodel, we construct a heterogeneous graph $\mathcal{G} = (\mathcal{V},\mathcal{E})$ to model the dual-community structure hidden within the videos, as showed in Figure~\ref{fig:main}. The node set $\mathcal{V}$ consists of three types of nodes: video, uploader, and event, formally expressed as $\mathcal{V}=\mathcal{V}_\text{video}\cup\mathcal{V}_\text{uploader}\cup\mathcal{V}_\text{event}$. Each video $x_i$ is represented as a node $v_i \in \mathcal{V}_\text{video}$, with its node embedding initialized using the content representation $\textbf{x}_{\textbf{v},i}^{0}$ extracted from the multi-modal information. In the following sections, we will explain how communities are formed through the construction of uploader and event nodes, as well as the creation of heterogeneous edges $\mathcal{E}$ that connect these nodes.

\subsubsection{Uploader community construction}
In \themodel, we construct the uploader community to model the implicit relationships between video creators based on their profile similarities. This is motivated by the idea that, uploaders often form communities based on shared interests, similar content creation patterns, and at times, coordinated malicious intent. Content from similar or grouped uploaders may exhibit clustering behavior that amplifies misleading narratives.

To capture this, we introduce uploader nodes into the graph $\mathcal{G}$, denoted as $\mathcal{V}_\text{uploader}$, and establish edges between uploader nodes and video nodes in $\mathcal{V}_\text{video}$. These edges represent the relationship between an uploader and the videos they submit. Mathematically, for each video $x_i$, we connect the uploader node $u_j\in\mathcal{V}_\text{uploader}$ with the corresponding video node $v_i \in \mathcal{V}_\text{video}$ through an edge $(u_j,v_i)\in \mathcal{E}_{u-v}$, representing the creator-submission relationship.

\textbf{Uploader profile feature representation} The uploader’s profile information $x_i^u$ plays a critical role in determining their position within the uploader community. To create a meaningful representation for the uploader node embedding, we start by transforming raw profile data into a structured input for feature extraction. Specifically, We construct input text structures that prioritize verification status and verification details, followed by the author introduction, and finally supplementary metrics like fan count, subscriptions, likes, etc. An example of the constructed input\footnote{Modified for illustration}:\\

\begin{mybox}
\textit{\underline{Primary information}: The author is a verified institutional uploader, and the introduction of the institutional verification is ``the official account of the Junior Reporter Department of Handan Daily''.\\
\vspace{1mm}
\underline{Secondary information}: The author introduction is: ``I am a Junior Reporter of Handan Daily! Committed to recording every activity of junior reporters and the talent display of junior reporters, leading you to enter a world of childlike innocence here! ''\\
\vspace{1mm}
\underline{Supplementary information}: 432,469 likes; 7,280 followers; 1,475 videos; 90 subscriptions; location: Hebei.}
\end{mybox}

We embed each uploader profile $x_i^u$ via an LLM, given that LLMs have exhibited strong language comprehension capabilities. Formally, the initial embedding of the uploader node $u_j$ associated with the video $v_i$ is generated as:
\begin{equation}
    \textbf{x}_{\textbf{u},j}^{0} = \mathbf{W}_u\text{LLM}(p(x_i^u))\quad \forall u_j\in \mathcal{V}_\text{uploader},(u_j,v_i)\in\mathcal{E}_{u-v}
\end{equation}
where $p(x_i^u)$ represents the process of constructing the structured text input for the uploader profile, $\mathbf{W}_u$ is a learnable projection matrix.

\textbf{Uploader community clustering} 
To build the uploader community, we employ a clustering approach to group uploaders with similar profiles. Specifically, we apply the K-means clustering algorithm to the set of uploader embeddings $\{\textbf{x}_{\textbf{u},i}^{0}\}$, partitioning uploaders into $K$ clusters. Each cluster represents a community of similar uploaders. The cluster centers, denoted as $\mathcal{C}_k$, serve as the representative nodes for each community. Formally, for each uploader $u_i$, we assign them to a cluster $k_i$ by:
\begin{equation}
    k_i =\arg \min_k \| \textbf{x}_{\textbf{u},i}^{0} - \mathcal{C}_k\|_2
\end{equation}
After clustering, we add each cluster center $\mathcal{C}_k$ as an uploader node $u_{\mathcal{C}_k}$ into the graph. Each uploader $u_i$ is then connected to its respective cluster center node $u_{\mathcal{C}_{k_i}}$ via an edge $(u_i, u_{\mathcal{C}_{k_i}}) \in \mathcal{E}_{u-u}$. Each cluster  center node is initialized as the mean of its member uploader embeddings.

Through clustering and connecting uploader nodes to their respective cluster centers, we form the uploader community. The edges between uploader nodes and cluster centers materialize the implicit communities in the graph structure, encoding profile similarities explicitly. 

\subsubsection{Event community construction}
In addition to the uploader community, public events also play a crucial role in gathering multiple videos, with different events exhibiting distinct patterns of misinformation dissemination. To capture this, we introduce the event community. 
Specially, we leverage each video's event information $x_i^e$--a descriptive tag (e.g., ``sleeper coach falling off bridge with casualties'') that identifies the public event to which the short video pertains.

We construct the event node set $\mathcal{V}_\text{event}$ by collecting all distinct events mentioned in the dataset $D$. Each video node $v_i \in \mathcal{V}_\text{video}$ is connected to its corresponding event node $e_l \in \mathcal{V}_\text{event}$ via an edge $(v_i, e_l) \in \mathcal{E}_{v-e}$, based on the video’s event annotation $x_i^e$. The initial embedding of each event node is obtained using an LLM:
\begin{equation}
    \textbf{x}_{\textbf{e},l}^{0}=\mathbf{W}_e\text{LLM}(x_i^e)\quad \forall e_l\in\mathcal{V}_\text{event},(v_i,e_l)\in\mathcal{E}_{v-e}
\end{equation}

\textbf{Event community relationships}
Besides the affiliation relationship established by $\mathcal{E}_{v-e}$, we introduce further edge constructions to model the interconnectedness within the event community. First, we connect video nodes associated with the same event, forming a network of related content under that event. These edges offer a broader perspective on related materials, facilitating the identification of coordinated misinformation campaigns.
\begin{equation}
    \mathcal{E}_{v-v} = \{(v_i,v_j)|(v_i,e_l)\in\mathcal{E}_{v-e},(v_j,e_l)\in\mathcal{E}_{v-e}\}
\end{equation}

Additionally, similar events tend to propagate false information in similar ways, while distinct events may reveal different patterns. To capture the relationship between events, we calculate the cosine similarity between the event node embeddings. If the similarity exceeds a threshold $\tau$, we establish an edge between the corresponding event nodes:
\begin{equation}
\mathcal{E}_{e-e} = \{(e_m,e_n)|\text{Sim}(\textbf{x}_{\textbf{e},m}^{0},\textbf{x}_{\textbf{e},n}^{0})\ge\tau\}  
\end{equation}

As a summary, we have constructed the dual-community graph $\mathcal{G}$, where the edge set $\mathcal{E}$ includes the relationships $\mathcal{E}_{u-v}$, $\mathcal{E}_{u-u}$, $\mathcal{E}_{v-e}$, $\mathcal{E}_{v-v}$, and $\mathcal{E}_{e-e}$, capturing the interactions within video, uploader and event.

\subsection{Time-Sensitive Heterogeneous Information Propagation}
Temporal dynamics are a crucial signal in misinformation detection, as fake content often emerges in bursts around major events~\cite{sharma2021identifying}. To capture such patterns, we extend the heterogeneous graph attention network (HGAT) by incorporating time-aware message passing over our dual-community graph $\mathcal{G}$.

We assign timestamps to video nodes by upload time, and to event nodes using the earliest upload time among associated videos. Uploader nodes are treated as time-invariant with no timestamp. For an edge between nodes $i$ and $j$, we define the time gap $\Delta t_{ij} = t_j - t_i$, and encode it as:
\begin{equation}
    \mathbf{t}_{ij} = 
    \begin{cases} 
    \operatorname{SinEnc}(\Delta t_{ij}), & \text{if both } i, j \text{ have timestamps} \\
    \mathbf{0}, & \text{otherwise}
    \end{cases}
\end{equation}
Here, $\operatorname{SinEnc}(\cdot)$ denotes a sinusoidal encoding function, mapping scalar time differences into fixed-length vectors, similar to positional encodings in Transformers~\cite{vaswani2017attention}.
The encoded time difference $\mathbf{t}_{ij}$ is integrated into the graph attention mechanism. At each layer $l$, node $i$ aggregates messages from its neighbors across all relation types $r \in \mathcal{R}$. The updated representation is given by:
\begin{equation}
\textbf{x}_{\mathbf{*},i}^{l+1} = \sum_{r\in\mathcal R}\sum_{j\in\mathcal{N}_i^r}\alpha_{i,j}^r\mathbf{\Theta}^{l}_{r,t}[\mathbf{x}_{\mathbf{*},j}^{l}
    \| \mathbf{t}_{ij}]
\end{equation}
\begin{equation}
\alpha_{ij}^r = \frac{\exp\left(\sigma\left(
[\mathbf{a}^{l}_{r,s}\|\mathbf{a}^{l}_{r,t}]^\text{T}[\textbf{x}_{\mathbf{*},i}^{l}\|\textbf{x}_{\mathbf{*},j}^{l}] + \mathbf{a}_{r,d}^{l\,\text{T}}\mathbf{t}_{ij}
\right)\right)}
{\sum_{k \in \mathcal{N}_i^r} \exp\left(\sigma\left(
[\mathbf{a}^{l}_{r,s}\|\mathbf{a}^{l}_{r,t}]^\text{T}[\textbf{x}_{\mathbf{*},i}^{l}\|\textbf{x}_{\mathbf{*},k}^{l}] + \mathbf{a}_{r,d}^{l\,\text{T}}\mathbf{t}_{ik}
\right)\right)}
\end{equation}
where $\mathbf{a}, \mathbf{\Theta}$ are learnable parameters, $\sigma$ is the LeakyReLU activation function. After $L$ layers of propagation, we obtain the final node representations $\mathbf{x}_{*,i}^{L}$, which integrate heterogeneous structural, semantic, and temporal signals.

\begin{table*}[tb]
  \centering
  \resizebox{\textwidth}{!}{
  \setlength{\tabcolsep}{10.8pt}
    \begin{tabular}{lcccccccc}
    \toprule[1.4pt]
    \multirow{2}{*}{\textbf{Model}} & \multicolumn{4}{c}{\textit{\textbf{Event-Level Split}}} & \multicolumn{4}{c}{\textit{\textbf{Temporal Split}}} \\
    \cmidrule{2-9} 
    & \textbf{Acc.} & \textbf{F1} & \textbf{Prec.} & \textbf{Rec.} & \textbf{Acc.} & \textbf{F1} & \textbf{Prec.} & \textbf{Rec.} \\
    \midrule[1.1pt]
    Bert & 77.30 & 77.27 & 77.43 & 77.30 & 80.81 & 80.32 & 80.80 & 80.07 \\
    FANVM & 75.49 & 75.46 & 75.71 & 75.53 & 82.66 & 82.21 & 82.72 & 81.94 \\
    SVFEND & 79.31 & 79.24 & 79.62 & 79.31 & 81.05 & 81.02 & 81.24 & 81.05 \\
    \midrule[0.8pt]
    FakingRecipe & 79.60 & 79.59 & 79.67 & 79.60 & 84.69 & 84.30 & 84.80 & 84.01\\
    MMVD & 82.64 & 82.63 & 82.63 & 82.73 & - & - & - & -\\
    MMSFD & 81.83 & 81.81 & 82.02 & 81.84 & - & - & - & - \\
    SVFEND-NEED & 84.62 & 84.61 & 84.81 & 84.64 & 89.67 & 89.37 & 90.16 & 88.97\\
    
    OpEvFake & - & - & - & - & 88.01 & 87.80 & 87.90 & 87.71 \\
    ExMRD & 80.48 & 80.46 & 80.67 & 80.51 & 86.90 & 86.52 & 87.31 & 86.13 \\
    GPT-4o-m & 67.10 & 67.08 & 67.15 & 67.21 & 66.42 & 65.88 & 65.90 & 65.87 \\
    LLaVA-OV & 58.14 & 54.76 & 60.44 & 57.51 & 57.54 & 50.71 & 61.57 & 55.94 \\
    \midrule[0.8pt]
     \rowcolor{gray!16}\textbf{Bert} + \textbf{\themodel} & 82.90 & 82.89 & 82.99 & 82.92 & 87.08 & 86.74 & 87.39 & 86.39 \\
     \rowcolor{gray!16}\textbf{FANVM} + \textbf{\themodel} & 82.44 & 82.42 & 82.52 & 82.43 & 86.16 & 85.49 & 87.92 & 84.79 \\
     \rowcolor{gray!16}\textbf{SVFEND} + \textbf{\themodel} & \textbf{85.50} & \textbf{85.49} & \textbf{85.66} & \textbf{85.51} & \textbf{90.96} & \textbf{90.71} & \textbf{91.47} & \textbf{90.03} \\
    \bottomrule[1.2pt]
    \end{tabular}%
    }
    \caption{Performance on the FakeSV dataset under event-level and temporal splits. \themodel is applied on top of different multi-modal feature extractors: BERT, FANVM, and SVFEND. It achieves improvements on both event and temporal accuracy: BERT (+5.60 / +6.27), FANVM (+6.95 / +3.50), and SVFEND (+6.19 / +9.91), respectively. All metrics are macro-averaged: Acc. = Accuracy, Prec. = Precision, Rec. = Recall, F1 = F1-score.}
  \label{tab:main}
\end{table*}

\subsection{Training and Inference}

We adopt a two-stage training strategy inspired by masked autoencoders~\cite{he2022masked}, consisting of a reconstruction-based pretraining phase followed by supervised classification.

\textbf{Pretraining via masked node reconstruction}
To enhance the representation learning of nodes, we randomly mask a proportion $q$ of nodes across all types and replace their input embeddings $\mathbf{x}_{*,i}^0$ with a learnable vector [MASK]. The model is then trained to reconstruct the original embeddings using a type-specific MLP decoder:
\begin{equation}
\mathcal{L}_\text{rec} = \sum_{z \in \{v,u,e\}} \sum_{i \in \mathcal{M}_z} \left\| \text{MLP}_{z}(\mathbf{x}_{\mathbf{z},i}^L) - \mathbf{x}_{\mathbf{z},i}^0 \right\|_2^2
\end{equation}
where $\mathcal{M}_{z}$ denotes the set of masked nodes of type $z$.

\textbf{Supervised classification}
After pretraining, we fine-tune the model on the labeled set $D_\text{labeled}$. A binary classifier is applied to the final representations of video nodes to predict whether the video is fake or real. The classification loss is a standard binary cross-entropy:
\begin{equation}
\mathcal{L}_\text{cls} = \sum_{x_i \in D_\text{labeled}} \text{BCE}(\text{MLP}_\text{cls}(\mathbf{x}_{\mathbf{v},i}^L),\ y_i)
\end{equation}
where $y_i \in \{0,1\}$ is the ground-truth label indicating the authenticity of video $x_i$.

During inference, we skip masking and directly use final-layer video node representations. Predictions for videos in $D_\text{unlabeled}$ are made using the trained classifier $\text{MLP}_\text{cls}$.

%% file: experiment.tex
\section{Experiment}
\subsection{Dataset}
We evaluate our model primarily on the FakeSV dataset~\cite{qi2023fakesv}, and additionally on FakeTT~\cite{bu2024fakingrecipe} for generalization analysis. 

FakeSV contains 3,624 short videos, with 1,810 labeled as fake and 1,814 as real, spanning 614 public events. Each video is accompanied by rich contextual information, including uploader profiles, video comments, and event tags. As evaluation metrics, we report accuracy and the macro-averaged precision, recall, and F1-score. We follow the same evaluation protocols proposed by \citet{qi2023fakesv}, considering both event-level and temporal split settings:
\begin{itemize}
    \item \textbf{Event-level split}: We conduct five-fold cross-validation, where in each fold, the dataset is divided by events in a 4:1 ratio for training and testing. This ensures that no videos from the same event appear in both sets, preventing information leakage.
    \item \textbf{Temporal split}: Videos are sorted by upload time and divided into 70\% training, 15\% validation, and 15\% test sets. This setup simulates a realistic scenario where the model predicts future videos based on past data.
\end{itemize}
Compared to FakeSV, FakeTT contains 1,991 videos (1,172 fake and 819 real) and provides weaker contextual signals, lacking video comments and offering limited uploader profile features (e.g., no follower counts). We apply the same evaluation settings to assess the robustness of our model in such low-context environments.

\subsection{Baselines}
To evaluate the effectiveness of \themodel, we compare it against 11 baseline approaches. Among them, BERT~\cite{devlin2019bert} serves as a single-modal baseline using only textual information. SVFEND~\cite{qi2023fakesv}, FANVM~\cite{choi2021using}, FakingRecipe~\cite{bu2024fakingrecipe}, MMVD~\cite{zeng2024mitigating}, and MMSFD~\cite{ren2024mmsfd} are multi-modal models designed specifically for fake short video detection, each adopting different fusion strategies.
NEED~\cite{qi2023two} constructs homogeneous video graphs within each event and incorporates debunking videos as external support. OpEvFake~\cite{zong2024unveiling} and ExMRD~\cite{hong2025following} integrate outputs from GPT-3.5 and GPT-4o-m~\cite{achiam2023gpt}, respectively, into their architectures via prompting.
For additional comparison, we also include zero-shot performance from GPT-4o-m and LLaVA-OV~\cite{li2024llava} as reference points.

\begin{table}[t]
\centering
\begin{tabular}{lcccc}
\toprule[1.2pt]
\textbf{Model} & \textbf{Acc.} & \textbf{F1} & \textbf{Prec.} & \textbf{Rec.} \\
\midrule[0.8pt]
\multicolumn{5}{l}{\small\textit{\textbf{Event-Level Split}}}\vspace{0.4mm} \\
SVFEND & 76.69 & 75.66 & 75.30 & 77.09 \\
\rowcolor{gray!16} SVFEND + \themodel & \textbf{79.20} & \textbf{78.61} & \textbf{78.59} & \textbf{81.01} \\
\midrule[0.5pt]
\multicolumn{5}{l}{\small\textit{\textbf{Temporal Split}}}\vspace{0.4mm} \\
SVFEND & 78.26 & 75.99 & 75.54 & 76.61 \\
\rowcolor{gray!16} SVFEND + \themodel & \textbf{84.62} & \textbf{82.81} & \textbf{82.51} & \textbf{83.14} \\
\bottomrule[1.2pt]
\end{tabular}
\caption{Performance on the FakeTT dataset under event-level and temporal splits.}
\label{tab:fakett}
\end{table}

\subsection{Implementation Details}
For uploader community construction, we group uploaders into $K=24$ clusters. For event communities, we retain only the top $p=2\%$ of pairwise connections based on similarity by applying the threshold $\tau$. To ensure efficiency, Qwen3-0.6B is used to encode uploader profiles and event descriptions. The hidden dimension of HGAT is set to 256, and the time difference vector $\mathbf{t}_{ij}$ is 16-dimensional, with time gaps measured in days. We stack four HGAT layers for message propagation.
During pretraining, $q=30\%$ of the nodes are randomly masked, and the model is trained for 50 epochs using reconstruction loss, followed by another 50 epochs of supervised fine-tuning with early stopping.
All experiments are conducted on a single Tesla-V100 GPU.

We apply \themodel on top of three base feature extractors: BERT, FANVM, and SVFEND. Each base model is first trained independently, and its intermediate features are used to initialize video node representations in \themodel. To ensure fairness and prevent data leakage, the base extractor and \themodel are trained separately but always on the same training data in each fold or temporal split.

For the FakeSV dataset, we adopt the same preprocessing procedure as \citet{qi2023fakesv}. For FakeTT, OCR features are extracted from the first frame of each video, and the comment is omitted due to missing data. We modify the original SVFEND pipeline accordingly to match this setup.

\subsection{Overall Performance}

As Table~\ref{tab:main} shows, \themodel consistently improves all three base extractors under event-level and temporal splits, outperforming multi-modal LLMs and prior fake short video detection models. Notably, \themodel also surpasses SVFEND-NEED, which leverages external debunking videos—whereas our method operates without relying on any external resources. Importantly, even though SVFEND already encodes uploader and event information, \themodel still yields substantial improvements: +6.19 accuracy on the event split and +9.91 on the temporal split. These results suggest that there remains rich structural and temporal information latent in uploader and event communities, which \themodel effectively captures through dual-community modeling and time-aware propagation. 

To test generalizability, we evaluate \themodel on the FakeTT dataset, which provides weaker contextual signals. As shown in Table~\ref{tab:fakett}, \themodel still brings consistent gains over the base model across both split settings, demonstrating its robustness in low-context environments.

\begin{table}[t]
\centering
\begin{tabular}{lcccc}
\toprule[1.4pt]
\textbf{Variant} & \textbf{Acc.} & \textbf{F1} & \textbf{Prec.} & \textbf{Rec.} \\
\midrule[0.8pt]
\textbf{Full Model} & \textbf{85.50} & \textbf{85.49} & \textbf{85.66} & \textbf{85.51} \\
\midrule
-- UploaderComm. & 82.94 & 82.94 & 82.95 & 82.94 \\
\arrayrulecolor{gray}\midrule[0.001pt]
-- EventComm. & 83.07 & 82.98 & 83.63 & 83.04 \\
\quad -- EventRel. & 84.81 & 84.79 & 84.96 & 84.80 \\
\quad -- VideoRel. & 83.97 & 83.96 & 84.05 & 83.97 \\
\midrule[0.001pt]
-- BothComm. & 80.80 & 80.80 & 80.82 & 80.81 \\
\arrayrulecolor{black}\midrule[0.5pt]
-- Pretrain & 84.22 & 84.16 & 84.58 & 84.20 \\
-- Time & 84.47 & 84.45 & 84.61 & 84.46\\
\bottomrule[1.2pt]
\end{tabular}
\caption{Ablation study on FakeSV (event-level split using SVFEND as base). “--” indicates the removal of the corresponding component; “Comm.” and “Rel.” denote community and relation, respectively.}
\label{tab:ablation}
\end{table}

\subsection{Ablation Study}
To assess the contribution of each component in \themodel, we conduct a detailed ablation study using SVFEND as the base extractor under the event-level split. Table~\ref{tab:ablation} summarizes the results. Our analysis focuses on the roles of community modeling (event/uploader), the reconstruction-based pretraining, and the time-aware propagation mechanism.

\textbf{Community Modeling} 
To test uploader community structure’s effect, we remove the community grouping process and directly use uploader nodes without graph construction. This variant, denoted as “{--UploaderComm.}”, causes a performance drop of 2.56 accuracy points, indicating the importance of capturing social structure among uploaders. 

For the event community, we distinguish two types of relations: video-to-video links within the same event ($\mathcal{E}_{v-v}$) and inter-event similarities ($\mathcal{E}_{e-e}$). Ablating each relation individually (“{--VideoRel.}” and “{--EventRel.}”) leads to moderate declines, confirming that both contribute complementary structural signals. Removing both yields the “{--EventComm.}” variant, which causes a larger drop of 2.43 in accuracy, highlighting the necessity of jointly modeling intra- and inter-event connections. 

Finally, “{--BothComm.}” variant disables both uploader and event communities entirely, resulting in a sharp 4.70-point drop, with performance only marginally above the base extractor. This underscores the strong complementary benefit of dual-community reasoning.

\textbf{Learning Configuration}
We further examine the impact of our learning setup. Removing the reconstruction-based pretraining (“{--Pretrain}”) results in a 1.28-point accuracy decline, suggesting that masked node reconstruction helps the model learn more transferable node representations. 

Disabling the time-aware propagation mechanism (“{--Time}”) degrades performance by 1.03 points, demonstrating that modeling temporal intervals between nodes is helpful for capturing the dynamics of misinformation spread.

\begin{figure}[t]
  \centering
  \begin{minipage}{0.49\linewidth}
    \centering
    \captionsetup{skip=2pt}
    \includegraphics[width=\linewidth]{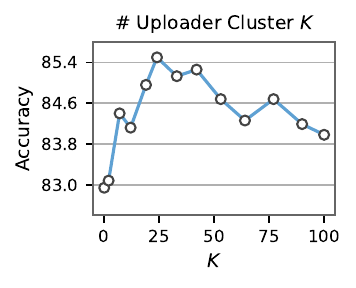}
    \caption{Effect of uploader cluster number $K$.}
    \label{fig:cluster}
  \end{minipage}
  \hfill
  \begin{minipage}{0.49\linewidth}
    \centering
    \captionsetup{skip=2pt}
    \includegraphics[width=\linewidth]{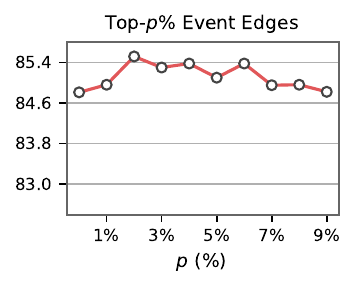}
    \caption{Effect of event similarity threshold.}
    \label{fig:threshold}
  \end{minipage}
\end{figure}

\begin{figure}[t]
  \centering
  \begin{minipage}{0.49\linewidth}
    \centering
    \captionsetup{skip=2pt}
    \includegraphics[width=\linewidth]{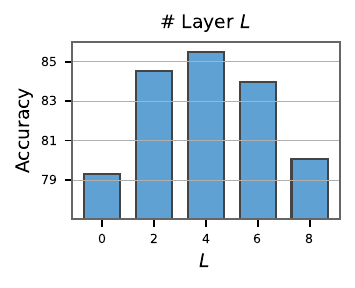}
    \caption{Effect of HGAT layer number $L$.}
    \label{fig:layer}
  \end{minipage}
  \hfill
  \begin{minipage}{0.49\linewidth}
    \centering
    \captionsetup{skip=2pt}
    \includegraphics[width=\linewidth]{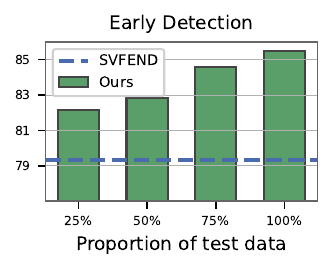}
    \caption{Performance of \themodel in early detection.}
    \label{fig:early}
  \end{minipage}
\end{figure}

\subsection{Hyper-parameter Analysis}
We analyze the sensitivity of \themodel to three key hyper-parameters: the number of uploader clusters $K$, the sparsity of event-event edges, and the number of HGAT layers $L$. We demonstrate result using SVFEND as the base extractor under the event-level split.

\paragraph{Uploader Cluster Number $K$}
We vary the number of uploader clusters from 0 (no uploader community) to 100. As shown in Figure~\ref{fig:cluster}, performance improves with finer clusters, peaking around $K=24$, then gradually declines as excessive fragmentation weakens community structure. This suggests that moderate clustering effectively captures social semantics, while over-partitioning leads to sparsity and noise.

\paragraph{Event Edge Sparsity}
To construct the event community, we retain only the top $p$\% of pairwise event similarities by applying a threshold $\tau$.  As shown in Figure~\ref{fig:threshold}, performance remains stable across a range of sparsity levels, but tends to degrade when the graph becomes overly dense. The best result is achieved when $p=2\%$, indicating that a sparse but semantically focused event graph is most effective. 

\paragraph{Number of HGAT Layers $L$}
We evaluate the impact of stacking 0 to 8 HGAT layers. As illustrated in Figure~\ref{fig:layer}, performance improves with depth up to $L=4$, beyond which additional layers yield diminishing or even negative returns. This highlights the importance of balancing propagation depth with over-smoothing risk, and supports our choice of using a four-layer HGAT in all experiments.

\begin{figure}[t]
    \centering
\includegraphics[width=\columnwidth]{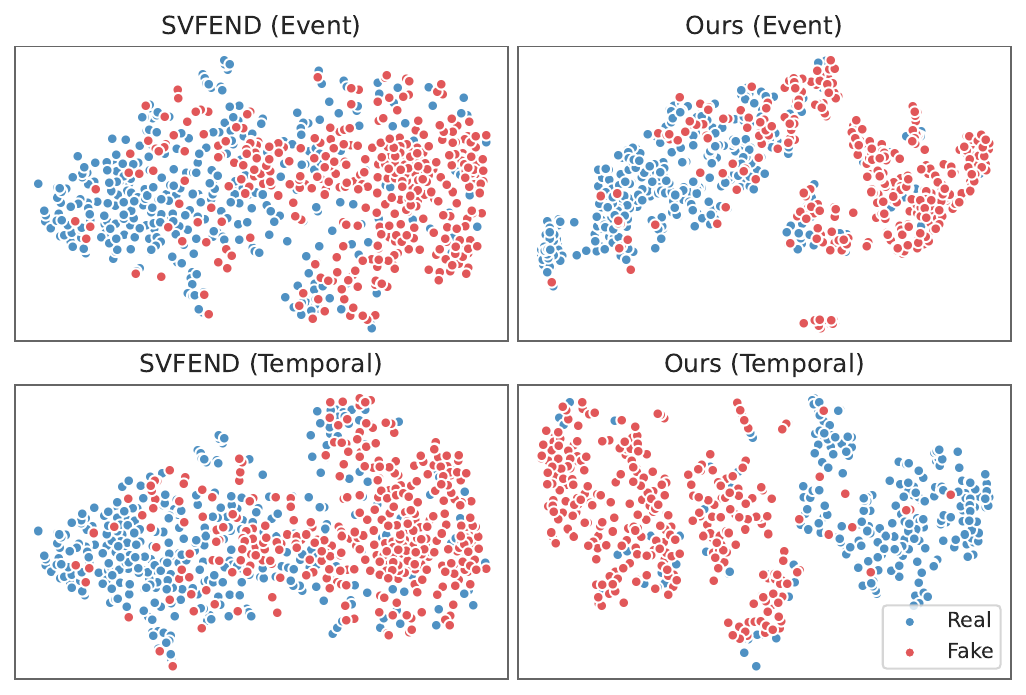}
    \caption{t-SNE visualizations of video features on the test set of FakeSV. Left: SVFEND; Right: SVFEND+\themodel. Top: event-level split. Bottom: temporal split.
}
    \label{fig:visual}
\end{figure}

\subsection{Early Detection}

Early detection is a critical aspect of evaluating a model's ability to identify fake news before it fully propagates. We assess how well \themodel performs when only a limited number of videos are available for a given event. Specifically, we simulate incomplete event contexts by selecting only 25\%, 50\%, 75\%, and 100\% of the videos associated with each event in the event-level test set. We then train and evaluate SVFEND+\themodel using the resulting incomplete dual-community graphs.
As shown in Figure~\ref{fig:early}, performance improves steadily as more event-related videos become available, enabling richer contextual reasoning. Even with just 25\% of the videos, \themodel still achieves a clear performance gain over the base SVFEND model, demonstrating its strong ability to reason under limited context.

\subsection{Visualization}
To further examine the representation quality, we visualize learned video node embeddings via t-SNE~\cite{maaten2008visualizing}. Figure~\ref{fig:visual} compares SVFEND and \themodel-enhanced feature distributions on the FakeSV test set under event-level and temporal splits. 
In both cases, SVFEND yields entangled feature spaces with limited separation between real and fake videos. In contrast, \themodel produces more compact and class-discriminative embeddings, with tighter clustering of real and fake instances.

%% file: conclusion.tex
\section{Conclusion}

In this work, we tackle the challenge of detecting fake news in short videos by modeling the community structures underlying uploader and event contexts.
We propose \themodel, a dual-community modeling framework that enhances existing video classifiers by incorporating uploader and event-level relationships. Our approach leverages a heterogeneous graph with time-aware message passing and self-supervised pretraining to capture both social semantics and temporal dynamics. Experiments on public benchmarks show consistent and significant performance gains across base video classifiers. 

%% file: main.bbl
\begin{thebibliography}{36}
\providecommand{\natexlab}[1]{#1}

\bibitem[{Achiam et~al.(2023)Achiam, Adler, Agarwal, Ahmad, Akkaya, Aleman, Almeida, Altenschmidt, Altman, Anadkat et~al.}]{achiam2023gpt}
Achiam, J.; Adler, S.; Agarwal, S.; Ahmad, L.; Akkaya, I.; Aleman, F.~L.; Almeida, D.; Altenschmidt, J.; Altman, S.; Anadkat, S.; et~al. 2023.
\newblock Gpt-4 technical report.
\newblock \emph{ArXiv preprint}, abs/2303.08774.

\bibitem[{A{\"\i}meur, Amri, and Brassard(2023)}]{aimeur2023fake}
A{\"\i}meur, E.; Amri, S.; and Brassard, G. 2023.
\newblock Fake news, disinformation and misinformation in social media: a review.
\newblock \emph{Social Network Analysis and Mining}, 13(1): 30.

\bibitem[{Bian et~al.(2020)Bian, Xiao, Xu, Zhao, Huang, Rong, and Huang}]{bian2020rumor}
Bian, T.; Xiao, X.; Xu, T.; Zhao, P.; Huang, W.; Rong, Y.; and Huang, J. 2020.
\newblock Rumor Detection on Social Media with Bi-Directional Graph Convolutional Networks.
\newblock In \emph{The Thirty-Fourth {AAAI} Conference on Artificial Intelligence, {AAAI} 2020, The Thirty-Second Innovative Applications of Artificial Intelligence Conference, {IAAI} 2020, The Tenth {AAAI} Symposium on Educational Advances in Artificial Intelligence, {EAAI} 2020, New York, NY, USA, February 7-12, 2020}, 549--556. {AAAI} Press.

\bibitem[{Bu et~al.(2024)Bu, Sheng, Cao, Qi, Wang, and Li}]{bu2024fakingrecipe}
Bu, Y.; Sheng, Q.; Cao, J.; Qi, P.; Wang, D.; and Li, J. 2024.
\newblock Fakingrecipe: Detecting fake news on short video platforms from the perspective of creative process.
\newblock In \emph{Proceedings of the 32nd ACM International Conference on Multimedia}, 1351--1360.

\bibitem[{Chen, Pan, and Zuo(2022)}]{chen2022tiktok}
Chen, Z.; Pan, S.; and Zuo, S. 2022.
\newblock TikTok and YouTube as sources of information on anal fissure: a comparative analysis.
\newblock \emph{Frontiers in Public Health}, 10: 1000338.

\bibitem[{Choi and Ko(2021)}]{choi2021using}
Choi, H.; and Ko, Y. 2021.
\newblock Using topic modeling and adversarial neural networks for fake news video detection.
\newblock In \emph{Proceedings of the 30th ACM international conference on information \& knowledge management}, 2950--2954.

\bibitem[{Cinelli et~al.(2021)Cinelli, De~Francisci~Morales, Galeazzi, Quattrociocchi, and Starnini}]{cinelli2021echo}
Cinelli, M.; De~Francisci~Morales, G.; Galeazzi, A.; Quattrociocchi, W.; and Starnini, M. 2021.
\newblock The echo chamber effect on social media.
\newblock \emph{Proceedings of the national academy of sciences}, 118(9): e2023301118.

\bibitem[{Devlin et~al.(2019)Devlin, Chang, Lee, and Toutanova}]{devlin2019bert}
Devlin, J.; Chang, M.-W.; Lee, K.; and Toutanova, K. 2019.
\newblock {BERT}: Pre-training of Deep Bidirectional Transformers for Language Understanding.
\newblock In \emph{Proceedings of the 2019 Conference of the North {A}merican Chapter of the Association for Computational Linguistics: Human Language Technologies, Volume 1 (Long and Short Papers)}, 4171--4186. Minneapolis, Minnesota: Association for Computational Linguistics.

\bibitem[{Er{\c{s}}ahin et~al.(2017)Er{\c{s}}ahin, Akta{\c{s}}, K{\i}l{\i}n{\c{c}}, and Akyol}]{ercsahin2017twitter}
Er{\c{s}}ahin, B.; Akta{\c{s}}, {\"O}.; K{\i}l{\i}n{\c{c}}, D.; and Akyol, C. 2017.
\newblock Twitter fake account detection.
\newblock In \emph{2017 international conference on computer science and engineering (UBMK)}, 388--392. IEEE.

\bibitem[{Francis(2024)}]{francis2024variation}
Francis, E. 2024.
\newblock Variation between credible and non-credible news across topics.
\newblock \emph{ArXiv preprint}, abs/2411.12458.

\bibitem[{Gong et~al.(2025)Gong, Li, Wu, Liu, Wu, and Wang}]{gong2025strive}
Gong, H.; Li, J.; Wu, J.; Liu, Q.; Wu, S.; and Wang, L. 2025.
\newblock STRIVE: Structured Reasoning for Self-Improvement in Claim Verification.
\newblock \emph{ArXiv preprint}, abs/2502.11959.

\bibitem[{Gong et~al.(2024{\natexlab{a}})Gong, Ma, Liu, Wu, and Wang}]{gong2024navigating}
Gong, H.; Ma, H.; Liu, Q.; Wu, S.; and Wang, L. 2024{\natexlab{a}}.
\newblock Navigating the noisy crowd: Finding key information for claim verification.
\newblock \emph{ArXiv preprint}, abs/2407.12425.

\bibitem[{Gong et~al.(2024{\natexlab{b}})Gong, Xu, Wu, Liu, and Wang}]{gong2024heterogeneous}
Gong, H.; Xu, W.; Wu, S.; Liu, Q.; and Wang, L. 2024{\natexlab{b}}.
\newblock Heterogeneous graph reasoning for fact checking over texts and tables.
\newblock In \emph{Proceedings of the AAAI Conference on Artificial Intelligence}, volume~38, 100--108.

\bibitem[{He et~al.(2022)He, Chen, Xie, Li, Doll{\'a}r, and Girshick}]{he2022masked}
He, K.; Chen, X.; Xie, S.; Li, Y.; Doll{\'a}r, P.; and Girshick, R. 2022.
\newblock Masked autoencoders are scalable vision learners.
\newblock In \emph{Proceedings of the IEEE/CVF conference on computer vision and pattern recognition}, 16000--16009.

\bibitem[{Hong et~al.(2025)Hong, Lang, Xu, Cheng, Zhong, and Zhou}]{hong2025following}
Hong, R.; Lang, J.; Xu, J.; Cheng, Z.; Zhong, T.; and Zhou, F. 2025.
\newblock Following clues, approaching the truth: Explainable micro-video rumor detection via chain-of-thought reasoning.
\newblock In \emph{Proceedings of the ACM on Web Conference 2025}, 4684--4698.

\bibitem[{Li et~al.(2024)Li, Zhang, Guo, Zhang, Li, Zhang, Zhang, Zhang, Li, Liu et~al.}]{li2024llava}
Li, B.; Zhang, Y.; Guo, D.; Zhang, R.; Li, F.; Zhang, H.; Zhang, K.; Zhang, P.; Li, Y.; Liu, Z.; et~al. 2024.
\newblock Llava-onevision: Easy visual task transfer.
\newblock \emph{ArXiv preprint}, abs/2408.03326.

\bibitem[{Li et~al.(2025)Li, Zhang, Xu, Li, Gao, and Wang}]{li2025learning}
Li, M.; Zhang, Y.; Xu, H.; Li, X.; Gao, C.; and Wang, Z. 2025.
\newblock Learning complex heterogeneous multimodal fake news via social latent network inference.
\newblock In \emph{Proceedings of the AAAI Conference on Artificial Intelligence}, volume~39, 433--441.

\bibitem[{Liu et~al.(2015)Liu, Zhan, Zhang, Sun, and Hui}]{liu2015events}
Liu, C.; Zhan, X.-X.; Zhang, Z.-K.; Sun, G.-Q.; and Hui, P.~M. 2015.
\newblock How events determine spreading patterns: information transmission via internal and external influences on social networks.
\newblock \emph{New Journal of Physics}, 17(11): 113045.

\bibitem[{Liu et~al.(2024)Liu, Wu, Wu, and Wang}]{liu2024out}
Liu, Q.; Wu, J.; Wu, S.; and Wang, L. 2024.
\newblock Out-of-distribution evidence-aware fake news detection via dual adversarial debiasing.
\newblock \emph{IEEE Transactions on Knowledge and Data Engineering}, 36(11): 6801--6813.

\bibitem[{Ma et~al.(2024)Ma, Zhang, Ding, Yang, Wu, and Fan}]{ma2024fake}
Ma, X.; Zhang, Y.; Ding, K.; Yang, J.; Wu, J.; and Fan, H. 2024.
\newblock On fake news detection with LLM enhanced semantics mining.
\newblock In \emph{Proceedings of the 2024 Conference on Empirical Methods in Natural Language Processing}, 508--521.

\bibitem[{Maaten and Hinton(2008)}]{maaten2008visualizing}
Maaten, L. v.~d.; and Hinton, G. 2008.
\newblock Visualizing data using t-SNE.
\newblock \emph{Journal of machine learning research}, 9(Nov): 2579--2605.

\bibitem[{Qi et~al.(2023{\natexlab{a}})Qi, Bu, Cao, Ji, Shui, Xiao, Wang, and Chua}]{qi2023fakesv}
Qi, P.; Bu, Y.; Cao, J.; Ji, W.; Shui, R.; Xiao, J.; Wang, D.; and Chua, T.-S. 2023{\natexlab{a}}.
\newblock Fakesv: A multimodal benchmark with rich social context for fake news detection on short video platforms.
\newblock In \emph{Proceedings of the AAAI Conference on Artificial Intelligence}, volume~37, 14444--14452.

\bibitem[{Qi et~al.(2023{\natexlab{b}})Qi, Zhao, Shen, Ji, Cao, and Chua}]{qi2023two}
Qi, P.; Zhao, Y.; Shen, Y.; Ji, W.; Cao, J.; and Chua, T.-S. 2023{\natexlab{b}}.
\newblock Two Heads Are Better Than One: Improving Fake News Video Detection by Correlating with Neighbors.
\newblock In \emph{Findings of the Association for Computational Linguistics: ACL 2023}, 11947--11959.

\bibitem[{Ren et~al.(2024)Ren, Liu, Zhu, Bing, Ma, and Wang}]{ren2024mmsfd}
Ren, S.; Liu, Y.; Zhu, Y.; Bing, W.; Ma, H.; and Wang, W. 2024.
\newblock MMSFD: Multi-grained and Multi-modal Fusion for Short Video Fake News Detection.
\newblock In \emph{2024 7th International Conference on Data Science and Information Technology (DSIT)}, 1--11. IEEE.

\bibitem[{Sharma et~al.(2021)Sharma, Zhang, Ferrara, and Liu}]{sharma2021identifying}
Sharma, K.; Zhang, Y.; Ferrara, E.; and Liu, Y. 2021.
\newblock Identifying coordinated accounts on social media through hidden influence and group behaviours.
\newblock In \emph{Proceedings of the 27th ACM SIGKDD conference on knowledge discovery \& data mining}, 1441--1451.

\bibitem[{Shu et~al.(2017)Shu, Sliva, Wang, Tang, and Liu}]{shu2017fake}
Shu, K.; Sliva, A.; Wang, S.; Tang, J.; and Liu, H. 2017.
\newblock Fake news detection on social media: A data mining perspective.
\newblock \emph{ACM SIGKDD explorations newsletter}, 19(1): 22--36.

\bibitem[{Vaswani et~al.(2017)Vaswani, Shazeer, Parmar, Uszkoreit, Jones, Gomez, Kaiser, and Polosukhin}]{vaswani2017attention}
Vaswani, A.; Shazeer, N.; Parmar, N.; Uszkoreit, J.; Jones, L.; Gomez, A.~N.; Kaiser, L.; and Polosukhin, I. 2017.
\newblock Attention is All you Need.
\newblock In Guyon, I.; von Luxburg, U.; Bengio, S.; Wallach, H.~M.; Fergus, R.; Vishwanathan, S. V.~N.; and Garnett, R., eds., \emph{Advances in Neural Information Processing Systems 30: Annual Conference on Neural Information Processing Systems 2017, December 4-9, 2017, Long Beach, CA, {USA}}, 5998--6008.

\bibitem[{Violot et~al.(2024)Violot, Elmas, Bilogrevic, and Humbert}]{violot2024shorts}
Violot, C.; Elmas, T.; Bilogrevic, I.; and Humbert, M. 2024.
\newblock Shorts vs. regular videos on YouTube: A comparative analysis of user engagement and content creation trends.
\newblock In \emph{Proceedings of the 16th ACM Web Science Conference}, 213--223.

\bibitem[{Wu et~al.(2024)Wu, Lin, Cao, and Lin}]{wu2024interpretable}
Wu, K.; Lin, Y.; Cao, D.; and Lin, D. 2024.
\newblock Interpretable short video rumor detection based on modality tampering.
\newblock In \emph{Proceedings of the 2024 Joint International Conference on Computational Linguistics, Language Resources and Evaluation (LREC-COLING 2024)}, 9180--9189.

\bibitem[{Zeng et~al.(2024)Zeng, Luo, Kong, Liu, Guo, Yang, Ma, and Zhao}]{zeng2024mitigating}
Zeng, Z.; Luo, M.; Kong, X.; Liu, H.; Guo, H.; Yang, H.; Ma, Z.; and Zhao, X. 2024.
\newblock Mitigating World Biases: A Multimodal Multi-View Debiasing Framework for Fake News Video Detection.
\newblock In \emph{Proceedings of the 32nd ACM International Conference on Multimedia}, 6492--6500.

\bibitem[{Zhang et~al.(2024{\natexlab{a}})Zhang, Gong, Liu, Wu, and Wang}]{zhang2024breaking}
Zhang, M.; Gong, H.; Liu, Q.; Wu, S.; and Wang, L. 2024{\natexlab{a}}.
\newblock Breaking event rumor detection via stance-separated multi-agent debate.
\newblock \emph{ArXiv preprint}, abs/2412.04859.

\bibitem[{Zhang et~al.(2024{\natexlab{b}})Zhang, Lv, Jia, Yun, Miao, Mao, and Wu}]{zhang2024gbca}
Zhang, Z.; Lv, Q.; Jia, X.; Yun, W.; Miao, G.; Mao, Z.; and Wu, G. 2024{\natexlab{b}}.
\newblock GBCA: Graph Convolution Network and BERT combined with Co-Attention for fake news detection.
\newblock \emph{Pattern Recognition Letters}, 180: 26--32.

\bibitem[{Zhao et~al.(2017)Zhao, Yao, Li, Song, and Lee}]{zhao2017meta}
Zhao, H.; Yao, Q.; Li, J.; Song, Y.; and Lee, D.~L. 2017.
\newblock Meta-Graph Based Recommendation Fusion over Heterogeneous Information Networks.
\newblock In \emph{Proceedings of the 23rd {ACM} {SIGKDD} International Conference on Knowledge Discovery and Data Mining, Halifax, NS, Canada, August 13 - 17, 2017}, 635--644. {ACM}.

\bibitem[{Zhong et~al.(2024)Zhong, Xiao, Xu, and Cheng}]{zhong2024vmid}
Zhong, W.; Xiao, Y.; Xu, M.; and Cheng, X. 2024.
\newblock VMID: A Multimodal Fusion LLM Framework for Detecting and Identifying Misinformation of Short Videos.
\newblock \emph{ArXiv preprint}, abs/2411.10032.

\bibitem[{Zhu et~al.(2024)Zhu, Gao, Yin, Li, and Kurths}]{zhu2024propagation}
Zhu, J.; Gao, C.; Yin, Z.; Li, X.; and Kurths, J. 2024.
\newblock Propagation Structure-Aware Graph Transformer for Robust and Interpretable Fake News Detection.
\newblock In \emph{Proceedings of the 30th ACM SIGKDD Conference on Knowledge Discovery and Data Mining}, KDD '24, 4652–4663. New York, NY, USA: Association for Computing Machinery.
\newblock ISBN 9798400704901.

\bibitem[{Zong et~al.(2024)Zong, Zhou, Lin, Liu, Zhang, and Xu}]{zong2024unveiling}
Zong, L.; Zhou, J.; Lin, W.; Liu, X.; Zhang, X.; and Xu, B. 2024.
\newblock Unveiling opinion evolution via prompting and diffusion for short video fake news detection.
\newblock In \emph{Findings of the Association for Computational Linguistics ACL 2024}, 10817--10826.

\end{thebibliography}
